\documentstyle[epsfig,prd,twocolumn,aps]{revtex} 
\begin{document}
%\draft         
\preprint{\vbox{\hbox{IFT--P.076/97}
		\hbox{FTUV/97-55}
		\hbox{IFIC/97-86}
		\hbox{hep-ph/9711446}
}}
\twocolumn[\hsize\textwidth\columnwidth\hsize\csname @twocolumnfalse\endcsname
\title{Search for Non--standard Higgs Boson in Diphoton Events 
       at $\boldmath{p\bar{p}}$ Collisions} 
\author{M.\ C.\ Gonzalez--Garcia$^{1,2}$, 
	S.\ M.\ Lietti$^1$ and S.\ F.\ Novaes$^1$}
\address{$^1$Instituto de F\'{\i}sica Te\'orica, 
             Universidade Estadual Paulista \\   
             Rua Pamplona 145,
             01405--900 S\~ao Paulo, Brazil} 
\address{$^2$Instituto de F\'{\i}sica Corpuscular -- IFIC/CSIC,
             Departament de F\'{\i}sica Te\`orica \\
             Universitat de Val\`encia, 46100 Burjassot, 
             Val\`encia, Spain}
\date{\today} 
\maketitle
\begin{abstract} 
We estimate the attainable limits on the coupling of a
non--standard Higgs boson to two photons taking into account the
data collected by the Fermilab  collaborations on diphoton
events.  We based our analysis on a general set of dimension--6
effective operators that give rise to anomalous couplings in the
bosonic sector of the Standard Model.  If the coefficients of all
``blind'' operators are of the same order of magnitude, bounds on
the anomalous triple vector--boson couplings can also be
obtained.
\end{abstract}
\vskip 2pc]

Events containing two photons plus large missing transverse
energy ($\gamma\gamma \not \!\! E_T$) represent an important
signature for some classes of supersymmetric models \cite{xer}.
Models that predict the existence of  light neutralinos
\cite{lit:neu} can give rise to this kind of event when the next
to lightest neutralino decays  $\tilde{\chi}_2^0 \to
\tilde{\chi}_1^0 \gamma$, where $\tilde{\chi}_1^0$ is the
lightest supersymmetric particle (LSP). When a light gravitino is
present \cite{lit:gra}, like in models with gauge--mediated low
energy supersymmetry breaking \cite{gau:med}, the lightest
neutralino is unstable and decays via $\tilde{\chi}_1^0 \to
\tilde{G} \gamma$, which also yield an event topology with two
photons together with missing energy, since the gravitino
($\tilde{G}$) escapes undetected. 

D\O ~Collaboration have reported a recent search for diphoton
events with large missing transverse energy in $p\bar{p}$
collisions at $\sqrt{s} = 1.8$ TeV \cite{prl:old,prl:new,eno}.
Their analysis indicates a good  agreement with the expectations
from the Standard Model (SM). In this way, D\O  ~Collaboration
were able to set limits on the production cross section
$\sigma(p\bar{p} \to \gamma\gamma \not \!\! E_T  + X)$, and
consequently, to establish an exclusion region in the
supersymmetry parameter space and lower bounds on the masses of
the lightest chargino and neutralino.

In this work, we point out that the experimental search for
$\gamma\gamma \not \!\! E_T$ events is also able to constraint
new physics in the bosonic sector of the SM. For instance,
associated Higgs--$Z$ boson production, with the subsequent decay
of the Higgs into two photons and the $Z$ going to neutrinos, can
yield this signature. In the SM, the decay width $H \to \gamma
\gamma$ is very small since it occurs just at one--loop level
\cite{h:gg}.  However, the existence of new interactions can
enhance this width  in a significant way. 

We can describe the deviations of the SM predictions for the
couplings in the bosonic sector via effective Lagrangians
\cite{classical,linear,drghm,hisz}. The new couplings among light
states are described by anomalous effective operators
representing residual interactions, after the heavy degrees of
freedom are integrated out. A complete set of eleven $C$ and $P$
conserving and $SU_L(2) \times U_Y(1)$ invariant operators can be
found  in Refs. \cite{linear,drghm,hisz}. The dimension--6
operators that alter the $HVV$ couplings, like $HWW$, $HZZ$,
$H\gamma\gamma$ and $HZ\gamma$, can be written in terms of the
Higgs doublet ($\Phi$) as 
\begin{eqnarray}
{\cal L}_{\text{eff}} &=& 
f_{WW} \Phi^{\dagger} \hat{W}_{\mu \nu} \hat{W}^{\mu \nu} \Phi + 
f_{BB} \Phi^{\dagger} \hat{B}_{\mu \nu} \hat{B}^{\mu \nu} \Phi 
\\ \label{lag}
&+& f_W (D_{\mu} \Phi)^{\dagger} \hat{W}^{\mu \nu} (D_{\nu} \Phi) 
+   f_B (D_{\mu} \Phi)^{\dagger} \hat{B}^{\mu \nu} (D_{\nu} \Phi) 
\nonumber
\end{eqnarray}
where $\hat{B}_{\mu\nu} = i (g'/2) B_{\mu \nu}$, and
$\hat{W}_{\mu \nu} = i (g/2) \sigma^a W^a_{\mu \nu}$, with
$B_{\mu \nu}$ and $ W^a_{\mu \nu}$ being the field strength
tensors of the $U(1)$ and $SU(2)$ gauge fields respectively.
Other possible operators like $\Phi^{\dagger}\hat{B}_{\mu
\nu}\hat{W}^{\mu \nu}\Phi$ (not ``blind'' operators) contribute
to gauge--boson two--point functions at tree level and are
strongly constrained.  The first two operators appearing in Eq.\
(\ref{lag}) do not modify the $WW\gamma$ and $WWZ$ tree--point
couplings, while the operators ${\cal O}_{W}$ and ${\cal O}_{B}$
generate both Higgs--vector boson  and self--vector--bosons
anomalous couplings. Therefore, the linearly realized effective
Lagrangians relate the modifications in the Higgs couplings to
those  in the vector boson vertex \cite{linear,drghm,hisz,hsz}.
It is important to notice that the coefficient of the operators
${\cal O}_{WW}$ and ${\cal O}_{BB}$ cannot be constrained by the
$W^+W^-$ production at LEP2, since they do not generate anomalous
triple gauge boson couplings. They can only be studied in
processes involving the Higgs boson in electron--positron
\cite{hsz,epem,ggg:bbg} or hadronic collisions \cite{prl}. 

We examine here the production of anomalously coupled Higgs boson
at Fermilab Tevatron $p\bar{p}$ collider. In particular, we
concentrate on the signature $\gamma\gamma \not \!\! E_T$ which
can originate from the reactions, 
\begin{eqnarray}
p \bar{p} & \to & Z (\to \nu  \bar{\nu}) + H (\to \gamma \gamma) + X
\nonumber \\ 
p \bar{p} & \to & W (\to \ell \nu) + H (\to \gamma \gamma) + X
\label{zw}  
\end{eqnarray}
where in the latter case the charged lepton ($\ell = e, \mu$)
escapes undetected. 

We have computed the cross sections (\ref{zw}) taking into
account all electroweak subprocess $q \bar{q}^\prime \to \nu
\bar{\nu} (\ell \nu) \gamma \gamma$, with $\ell = e, \mu$. The
anomalous contributions coming from the Lagrangian (\ref{lag})
and the interference with the SM diagrams were consistently
included via modified Helas \cite{helas} subroutines.  For the
proton structure functions, we have employed the MRS (G) set
\cite{mrs} at the scale $Q^2 = \hat{s}$.

In order to compare our predictions with the data collected by
the D\O ~Collaboration, we have applied the same cuts of Ref.\
\cite{prl:new}. We required that one photon has transverse energy
$E_T^{\gamma_1} > 20$ GeV and the other $E_T^{\gamma_2} > 12$
GeV, each of them with pseudorapidity in the range $|\eta^\gamma| <
1.2$ or  $1.5 < |\eta^\gamma| < 2.0$. We further required that
$\not \!\!E_T > 25$ GeV.  For the $\ell \nu \gamma \gamma$
final state, we imposed that the charged lepton is outside the
covered region of the electromagnetic calorimeter and it escapes
undetected  ($|\eta_{e}|> 2$ or $1.1<|\eta_e|<1.5 $, $
|\eta_{\mu}|> 1$). After these cuts we find that 80\% to 90\% of
the signal comes from associated Higgs--$Z$ production while 10\%
to 20\% arrises from Higgs--$W$. We also include in our analysis
the particle identification and trigger efficiencies which vary
from 40\% to 70\% per photon \cite{fermilab}. We estimate the
total effect of these efficiencies to be 35\%. 

The main sources of background to this reaction \cite{prl:new}
arise from SM processes containing multijets, direct photon, $W +
\gamma$, $W + j$, $Z \to ee$ and $Z \to \tau\tau \to ee$ where
photons are misidentified and/or the missing energy is
mismeasured. The D\O ~Collaboration estimate the contribution of
all these backgrounds to yield $2.3 \pm 0.9$ events.  D\O
~Collaboration has observed 2 events that have passed the above
cuts in their data sample of $106.3 \pm 5.6$ pb$^{-1}$. The
invariant mass of the photon pair in these events are $50.4$, and
$264.3$ GeV \cite{eno}. 

\begin{figure}
\begin{center}
\mbox{\epsfig{file=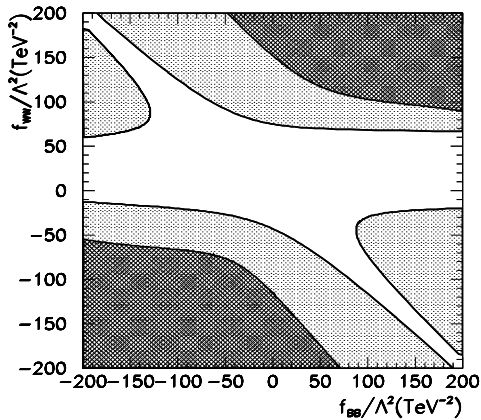,width=0.45\textwidth,height=0.3\textheight}}
\end{center} 
\caption{Excluded region at 95\% of CL in the $f_{WW} \times
f_{BB}$ plane, for an integrated luminosity of 100 pb$^{-1}$, and
for $M_H = 80 (140)$ GeV  [light shadow (dark shadow)].}
\label{fig:1} 
\end{figure}

In our analysis, we search for Higgs boson with mass in the range
$70 < M_H \lesssim 2 M_W$, since after the $W^+W^-$ threshold is
reached the diphoton branching ratio of Higgs is quite reduced.
Since no event with two--photon invariant mass in the range $70 <
M_{\gamma\gamma} \lesssim 2 M_W$ were observed, a $95\%$ CL in
the determination of the anomalous coefficient $f_i$, $i=WW, BB,
W, B$ of Eq.\ (\ref{lag}) is attained requiring 3 events coming
only from the anomalous contributions.

\begin{figure}
\begin{center}
\mbox{\epsfig{file=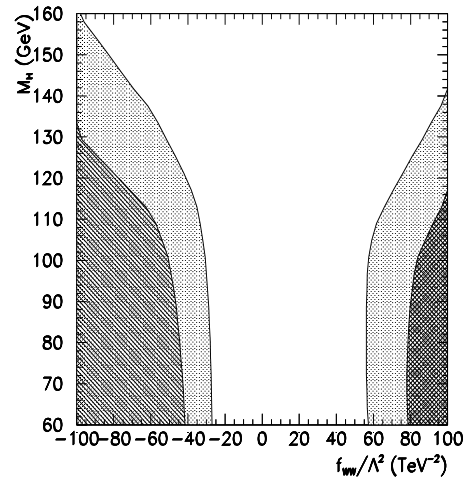,width=0.3\textwidth,height=0.3\textheight}}
\end{center} 
\caption{Excluded region $f_{WW} \times M_H$ plane for an
integrated luminosity of 100 pb$^{-1}$ at 64 \%  (95 \%)
CL  [light shadow (dark shadow)].}
\label{fig:2}
\end{figure}

In Fig.\ \ref{fig:1}, we present the exclusion region in the
$f_{WW}\times f_{BB}$ plane, when we assume that just these two
coefficients are different from zero. The clear (dark) shadow
represents the excluded region, at $95\%$ CL, for $M_H = 80
\, (140)$ GeV. We have used and integrated luminosity of 100
pb$^{-1}$ . Since the anomalous contribution to $H \to
\gamma\gamma$ width becomes zero for $f_{WW} = - f_{BB}$ a very
loose bound is obtained near this  axis. We should also notice
that the reactions (\ref{zw}) are more sensible to $f_{WW}$,
while the dependence on $f_{BB}$ is very weak. In Fig.\
\ref{fig:2}, we show the $f_{WW}$ values that can be excluded as
a function of the Higgs boson mass at $64$ \%  ($95\%$) CL. 

\begin{figure}
\begin{center}
\mbox{\epsfig{file=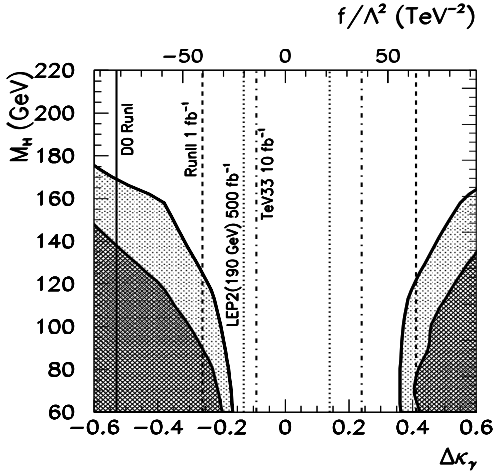,width=0.45\textwidth,height=0.3\textheight}}
\end{center} 
\caption{Excluded region in the $\Delta\kappa_\gamma \times M_H$
plane for an integrated luminosity of 100 pb$^{-1}$, and for  $f
\equiv f_{WW} = f_{BB}=$ $f_{W} = f_{B}$ ($f \equiv f_{WW} =
f_{BB} =$ $-f_{W} = -f_{B}$) [light shadow (dark shadow)]. The
vertical lines represent the present and future limits on
$\Delta\kappa_\gamma$ from different colliders.}
\label{fig:3}
\end{figure}

When we assume that all the coefficients of the Lagrangian
(\ref{lag}) have the same magnitude  the $H \to \gamma\gamma$
coupling becomes related to the triple vector boson coupling,
$WW\gamma$. Therefore, the limits obtained from Higgs production,
with the subsequent decay into two photons, is able to generate
an indirect bound on $\Delta \kappa_\gamma$
\cite{linear,drghm,hisz,hsz,prl}. In Fig.\ \ref{fig:3}, we
compare our indirect limit on $\Delta \kappa_\gamma$ with the
experimental limit of D\O ~Collaboration  from gauge boson pair
production \cite{fermilab} for $f \equiv f_{WW} = f_{BB} = f_{W}
= f_{B}$ (light shadow) and $f \equiv f_{WW} = f_{BB} = -f_{W} =
-f_{B}$ (dark shadow). We also display  the  expected bounds at
the upgraded Tevatron (Run II) and at TeV33, assuming 1 and 10
fb$^{-1}$ of integrated luminosity, respectively \cite{tevatron},
and the limit  that will be possible to extract from LEP II,
operating at 190 GeV with an integrated luminosity of 500
fb$^{-1}$ \cite{lep}.   We can see that, for $M_H \lesssim 170 \,
(140)$ GeV, the limit that can be  established at 95\% CL from
our analysis based on the present Tevatron luminosity is tighter
than the present limit coming from gauge boson production.  If
the result from the recent global fit to LEP, SLD, $p\bar{p}$,
and low energy data that favors a Higgs boson with mass  $M_H =
127 \stackrel{+127}{\scriptstyle{-72}}$ GeV \cite{ewwg} is not
substantially modified by the presence of the new operators, our
indirect limit on $\Delta\kappa_\gamma$ applies for the most
favoured Higgs masses. 

In conclusion, we have shown how to extract important information
on anomalous Higgs boson coupling from the analysis of
$\gamma\gamma \not \!\! E_T$ events in $p\bar{p}$ collisions. In
particular, we were able to establish limits on the coefficients
of general effective operators that give rise to the coupling
$H\gamma\gamma$. Since linearly realized effective Lagrangians
relate the modifications in the Higgs couplings to the ones
involving vector boson self--interaction, one can extract
indirect limits on the anomalous $WW\gamma$ coupling that are
competitive with the bounds from direct searches in gauge boson
production at  present and future collider experiments. 

\acknowledgments
The authors would like to thank O.\ J.\ P.\ \'Eboli for
suggesting this work. We also want to thank Sarah Eno for
providing us with information on the D\O ~data on two photons
plus missing transverse energy data. M.\ C.\ G--G is grateful to
the Instituto de F\'{\i}sica Te\'orica for its kind hospitality.
This work was supported by Funda\c{c}\~ao de Amparo \`a Pesquisa
do Estado de S\~ao Paulo (FAPESP), by DGICYT under grant
PB95-1077, by CICYT under grant AEN96--1718, and by Conselho
Nacional de Desenvolvimento Cient\'{\i}fico e Tecnol\'ogico
(CNPq).


\begin{references}
\bibitem{xer} For a recent review on the phenomenological aspects
of  supersymmetry see for instance: X.\ Tata, in the  Proceedings
of the IX Jorge Andr\'e Swieca Summer School: Particles and
Fields, S\~ao Paulo, Brazil, report UH-511-872-97, and
hep-ph/9706307.

\bibitem{lit:neu} S.\ Ambrosanio {\it et al.}, Phys.\ Rev.\
Lett.\  {\bf 76}, 3498 (1996); {\it idem} Phys.\ Rev.\ D{\bf 55}
1372 (1997).

\bibitem{lit:gra} S.\ Ambrosanio {\it et al.}, Phys.\ Rev.\ D{\bf
54},  5395 (1996); J.\ Kim {\it et al.}, ACT-11-97, and
hep-ph/9707331 

\bibitem{gau:med} P.\ Fayet, Phys.\ Lett.\ {\bf B70}, 461 (1977);
S.\ Dimopoulos {\it et al.}, Phys.\ Rev.\ Lett.\  {\bf 76}, 3494
(1996); H.\ Baer, M.\ Brhlik, C.\ Chen, and X.\  Tata, Phys.\
Rev.\ D{\bf 55} 4463 (1997).

\bibitem{prl:old} S.\ Abachi {\it et al.}, D\O ~Collaboration,
Phys.\ Rev.\ Lett.\ {\bf 78}, 2070 (1997). 

\bibitem{prl:new} B.\ Abbott {\it et al.}, D\O ~Collaboration,
FERMILAB-PUB-97-273-E, and hep-ex/9708005.

\bibitem{eno} See also the D\O ~Collaboration public Web page: 
\verb+http://www-d0.fnal.gov/public/new/analyses/gauge+
\verb+/welcome.html+.

\bibitem{h:gg} J.\ Ellis, M.\ K.\ Gaillard, D.\ V.\ Nanopoulos,
Nucl.\  Phys.\ {\bf B106}, 292 (1976); M.\ A.\ Shifman, A.\ I.\
Vainshtein, M.\  B.\ Voloshin, V.\ I.\ Zakharov, Sov.\ J.\ Nucl.\
Phys.\ {\bf 30}, 711  (1979).

\bibitem{classical}  K.\ Hagiwara, H.\ Hikasa, R.\ D.\ Peccei and
D.\ Zeppenfeld, Nucl.\ Phys.\ {\bf B282}, 253 (1987).

\bibitem{linear} C.\ J.\ C.\ Burguess and H.\ J.\ Schnitzer,
Nucl.\ Phys.\ {\bf B228}, 464 (1983); C.\ N.\ Leung, S.\ T.\ Love
and S.\ Rao, Z.\ Phys.\ {\bf 31}, 433 (1986);  W.\ Buchm\"uller
and D.\ Wyler, Nucl.\ Phys.\ {\bf B268}, 621 (1986).

\bibitem{drghm} A.\ De Rujula, M.\ B.\ Gavela, P.\ Hernandez and
E.\ Masso, Nucl.\ Phys.\ {\bf B384}, 3 (1992); A.\ De Rujula,
M.\ B.\ Gavela, O.\ Pene and F.\ J.\ Vegas,  Nucl.\ Phys.\ {\bf
B357}, 311 (1991).

\bibitem{hisz} K.\ Hagiwara, S.\ Ishihara, R.\ Szalapski and D.\
Zeppenfeld,  Phys.\ Lett.\ {\bf B283}, 353 (1992); {\it idem},
Phys.\ Rev.\ {\bf D48}, 2182 (1993); K.\ Hagiwara, T.\ Hatsukano,
S.\ Ishihara and  R.\ Szalapski, Nucl.\ Phys.\ {\bf B496}, 66
(1997).

\bibitem{hsz} K.\ Hagiwara, R.\ Szalapski and D.\ Zeppenfeld,
Phys.\ Lett.\ {\bf B318}, 155 (1993).

\bibitem{epem} K.\ Hagiwara, and M.\ L.\ Stong, Z.\ Phys.\ {\bf
C62},  99 (1994); B.\ Grzadkowski, and J.\ Wudka, Phys.\ Lett.\
{\bf B364},  49 (1995); G.\ J.\ Gounaris, J.\ Layssac and F.\ M.\
Renard,  Z.\ Phys.\ {\bf C65}, 245 (1995); G.\ J.\ Gounaris, F.\
M.\ Renard  and N.\ D.\ Vlachos, Nucl.\ Phys.\ {\bf B459}, 51
(1996).

\bibitem{ggg:bbg} S.\ M.\ Lietti, S.\ F.\ Novaes and R.\
Rosenfeld, Phys.\ Rev.\ {\bf D54}, 3266 (1996); F.\ de Campos,
S.\ M.\ Lietti, S.\ F.\ Novaes and R.\ Rosenfeld, Phys.\ Lett.\
{\bf B389}, 93 (1996); {\it idem}, Phys.\ Rev.\ {\bf D56}, 4384
(1997); S.\ M.\ Lietti and S.\ F.\ Novaes, hep--ph/9708443,
Phys.\ Lett.\ (in press).

\bibitem{prl} F.\ de Campos, M.\ C.\ Gonzalez--Garcia and S.\ F.\
Novaes, Phys.\ Rev.\ Lett.\ (in press).

\bibitem{helas} H.\ Murayama, I.\ Watanabe and K.\ Hagiwara, KEK
report 91-11 (unpublished).

\bibitem {mrs} A.\ D.\ Martin, W.\ J.\ Stirling, R.\ G.\ Roberts
Phys.\ Lett.\ {\bf B354}, 155 (1995).

\bibitem{fermilab} B.\ Abbott {\it et al.}, D\O  ~Collaboration,
Phys.\  Rev.\ Lett.\ {\bf 79}, 1441 (1997).

\bibitem{tevatron} D.\ Amidei {\it et al.}, {\it Future
Electroweak Physics at the Fermilab Tevatron: Report of the
TeV--2000 Study Group}, preprint FERMILAB-PUB-96-082 (1996).

\bibitem{lep} T.\ Barklow {\it et al.}, Summary of the Snowmass
Subgroup on Anomalous Gauge  Boson Couplings, to appear in the
{\it Proceedings of the 1996 DPF/DPB Summer Study on New
Directions in High-Energy Physics}, June 25 --- July 12 (1996),
Snowmass, CO, USA, and hep-ph/9611454.

\bibitem{ewwg} The LEP Collaborations ALEPH, DELPHI, L3, OPAL,
and the LEP Electroweak Working Group, contributions to the 28th
International Conference on High-energy Physics (ICHEP 96),
Warsaw, Poland, 1996, report CERN-PPE/96-183 (1996); {\it idem},
contributions to the 1997 Winter Conferences, LEPEWWG/97--01
(1997).

\end{references}
\end{document}